\documentclass[prb,preprint]{revtex4-1}

\usepackage{graphicx}
\usepackage{amssymb}
\usepackage{epstopdf}
\usepackage{amsmath}
\usepackage{bm}
\usepackage{algorithm}
\usepackage{algorithmic}
\usepackage{setspace}
\usepackage[raggedright]{subfigure}

\newcommand\refeq[1]{Eq. (\ref{#1})}
\newcommand\reffig[1]{Fig.~\ref{#1}}
\newcommand\of[1]{\left( #1 \right)}

\newcommand\vecbf[1]{{\bf #1}}
\newcommand\meskip{\hbox{\hskip 1cm}}

\begin{document}

\title{A Magnetic Velocity Verlet Method}

\author{A. Chambliss}
\email{chambgam@alumni.reed.edu}

\author{J. Franklin}
\email{jfrankli@reed.edu}
\affiliation{Department of Physics, Reed College, Portland, Oregon 97202,
USA}

\begin{abstract}
We discuss an extension of the velocity Verlet method that accurately approximates  the kinetic-energy-conserving charged particle motion that comes from magnetic forcing.  For a uniform magnetic field, the method is shown to conserve both particle kinetic energy and magnetic dipole moment better than midpoint Runge-Kutta.  We then use the magnetic velocity Verlet method to generate trapped particle trajectories, both in a cylindrical magnetic mirror machine setup, and for dipolar fields like the earth's magnetic field.  Finally, the method is used to compute an example of (single) mirror motion in the presence of a magnetic monopole field, where the trajectory can be described in closed form.
\end{abstract}

\maketitle

\section{Introduction}

The motion of particles under the influence of magnetic forces is complicated.  We know that a particle's speed does not change but that does not help us predict or visualize the twists and turns induced by non-uniform fields.  Our one concrete analytically tractable example is motion in a uniform magnetic field, hardly an advertisement for the exotic, spirograph-like trajectories that can appear in general field configurations.  Some particle trajectories can be described using special functions,\cite{APMOT} but these are not always accessible to students who are unfamiliar with those functions.

Absent a closed-form solution, there are various types of predictions one can make about the behavior of particles moving in non-trivial fields.  These are almost always either approximate or incomplete.  As an example of the latter, one of our favorites is Griffiths's Problem 5.43,\cite{GRIFFITHS} inviting students to show that a particle that starts at the center of a circular flux-free field region exits perpendicular to its boundary.  That demonstration is essentially an exercise in angular momentum conservation.  But what if the particle doesn't exit; what does it do instead?~\cite{FRANKLINAJP}  Another example, the ``magnetic mirror," in which particles are deflected from a region of converging magnetic field lines, is explored later in this article.  Beyond these almost qualitative descriptions, one must use either canned numerical methods,~\cite{VANALLEN, MAPLES,COMPUTERSTUDIES} or a homemade implementation of a method that can handle forces that depend on velocity in order to generate arbitrarily accurate approximations to particle trajectories.

One such approach, the Runge-Kutta method, while almost universally applicable to systems of ODEs, is not informed by the underlying physical problem that it solves.  In this paper we extend a different numerical integrator, the Verlet method, to include magnetic forces.  This update to the method, while known,~\cite{MDVERLET} is not widely used or taught at the undergraduate level.  Its derivation is already of interest to students thinking about the Lorentz force, and its ease of implementation, accuracy, and speed make it a desirable tool in a physicists' numerical toolbox.

After deriving the method, we demonstrate its energy conservation superiority over midpoint Runge-Kutta using a uniform magnetic field as a test case.  Then we proceed to both motivate and numerically generate trajectories for a magnetic mirror configuration.~\cite{JACKSON}  In particular, we use the magnetic moment, the ``first adiabatic invariant,"~\cite{JACKSON,GOLDSTEIN} to show that there can exist oscillatory motion along a field line.  Then we generate numerical trajectories that realize this oscillatory motion using the magnetic Verlet extension, and test the constancy of the kinetic energy and adiabatic invariant.  Having established the numerical existence of oscillatory trajectories in a field with cylindrical symmetry, we apply the method to a dipolar field to see if it can exhibit the oscillatory behavior observed in the cylindrical case.  For the magnetic field of a dipole, the equation of particle motion defines ``St\"ormer's problem," and we can use the magnetic Verlet extension to solve it numerically.  The dipole field can produce particle motion that spirals tightly along a field line while moving up and back along it and a nearby one.  As a final example, we demonstrate mirror motion for a particle moving in a magnetic monopole field, where we have an exact solution with which to compare.

\section{The Verlet Method}

The Verlet method generates approximate solutions to Newton's second law for position-dependent forces.  Its simplest variant, position Verlet,~\cite{VERLET} can be obtained quickly from Taylor expansion.  The expansion of a particle's position vector at time $t \pm \Delta t$ for small $\Delta t$ is
\begin{equation}
{\bf x}(t \pm \Delta t) \approx {\bf x}(t) \pm \dot{\bf x}(t) \Delta t + \frac{1}{2} \ddot{\bf x}(t) \Delta t^2 \pm \frac{1}{6} \dddot{\bf x}(t) \Delta t^3 + \frac{1}{24} \ddddot{\bf x}(t) \Delta t^4 \pm \ldots
\end{equation}
Adding together ${\bf x}(t+\Delta t)$ and ${\bf x}(t - \Delta t)$ in order to cancel the $\dot{\bf x}(t)$ term, we get
\begin{equation}\label{PVpre}
{\bf x}(t + \Delta t) + {\bf x}(t - \Delta t) = 2 {\bf x}(t) + \ddot{\bf x}(t) \Delta t^2 + O(\Delta t^4),
\end{equation}
where $O(\Delta t^4)$ means that there are errors of size $\Delta t^4$.  For a particle of mass $m$ moving under the influence of a force ${\bf F}({\bf x}, t)$, Newton's second law is $m \ddot{\bf x}(t) = {\bf F}({\bf x}(t),t)$, and we can write $\ddot{\bf x}(t)$ in terms of ${\bf x}(t)$.  Then the sum in~\refeq{PVpre} can be solved for ${\bf x}(t+\Delta t)$,
\begin{equation}\label{posVerlxup}
{\bf x}(t+\Delta t) = 2 {\bf x}(t) - {\bf x}(t - \Delta t) + \frac{1}{m} {\bf F}({\bf x}(t),t) \Delta t^2 + O(\Delta t^4).
\end{equation}
Viewed as an update method, this equation gives an approximation to the position of a particle at time $t + \Delta t$ based on its position at the current time $t$ and previous $t - \Delta t$, both of which are known.  The method is manifestly form-invariant under time-reversal.  Moving ${\bf x}(t+\Delta t)$ to the right hand side, and ${\bf x}(t - \Delta t)$ over to the left, we can run the dynamics backwards, from $t$ to $t - \Delta t$:
\begin{equation}
{\bf x}(t-\Delta t) = 2 {\bf x}(t) - {\bf x}(t + \Delta t) + \frac{1}{m} {\bf F}({\bf x}(t),t) \Delta t^2 + O(\Delta t^4),
\end{equation}
and this mimics the time-reversibility of Newton's second law itself.  The method is an example of a ``symplectic integrator," a class of numerical methods that preserve certain structural properties of the underlying Hamiltonian system.~\cite{SYMP}  Such methods do a good job conserving total energy, and it is this feature that recommends Verlet to our attention here.

If we wanted to test the energy conservation numerically, we would need access to the velocity at time $t$ in order to construct the kinetic energy.  There are a variety of ways to extract that information, but one of the more popular approaches is to re-order the position update in~\refeq{posVerlxup} into separate position and velocity updates.  There is nothing new here, just a rearrangement and labelling, leading to the velocity Verlet method defined by the updates:~\cite{AT} 
\begin{eqnarray}
{\bf x}(t + \Delta t) &=& {\bf x}(t) + {\bf v}(t)\Delta t  + \frac{1}{2 m} {\bf F}({\bf x}(t),t) \Delta t^2, \hbox{ and } \label{vvdefx}  \\
{\bf v}(t + \Delta t) &= & {\bf v}(t) + \frac{1}{2 m} \left[ {\bf F}({\bf x}(t),t) + {\bf F}({\bf x}(t + \Delta t), t+\Delta t)\right] \Delta t. \label{vvdefv}
\end{eqnarray}
The velocity approximation is less accurate than the position one, making the method as a whole similar in accuracy to midpoint Runge-Kutta,~\cite{FRANKLINCMP} which makes the same error in both position and velocity components.  In terms of timing, velocity Verlet requires two evaluations of the force at each update step, just as midpoint Runge-Kutta does.  For these reasons, we will compare the velocity Verlet method to the midpoint Runge-Kutta method when evaluating the former's numerical properties.

The velocity Verlet method is easily applied to conservative forces that depend only on position, but unlike Runge-Kutta methods, it is more difficult to introduce velocity-dependent forces like damping.  This difficulty is clear from the form of the updates in~\refeq{vvdefx} and~\refeq{vvdefv}:  while the updated positions are known going into the velocity update, if there were a force that depended on velocity, ${\bf F}({\bf x}, {\bf v}, t)$,~\refeq{vvdefv} would become:
\begin{equation}\label{vimp}
{\bf v}(t + \Delta t) = {\bf v}(t) + \frac{1}{2 m} \left[ {\bf F}({\bf x}(t),{\bf v}(t), t) + {\bf F}({\bf x}(t + \Delta t), {\bf v}(t + \Delta t),  t+\Delta t)\right] \Delta t,
\end{equation}
and it is not clear how to solve for ${\bf v}(t + \Delta t)$ since it appears on both the left and right-hand sides of this equation.  For a pure magnetic force, it is possible to isolate ${\bf v}(t + \Delta t)$ algebraically, as will be shown in the next section.

\section{Magnetic Velocity Verlet Method}
Consider a particle with charge $q$ moving in a given magnetic field ${\bf B}({\bf x})$, so that the Lorentz force is ${\bf F}({\bf x}, {\bf v}) = q {\bf v} \times {\bf B}({\bf x})$.  The velocity update in~\refeq{vimp} can be written 
\begin{equation}\label{LorentzUp}
{\bf v}(t + \Delta t) = {\bf v}(t) + \frac{q \Delta t}{2 m} \left[ {\bf v}(t) \times {\bf B}({\bf x}(t)) +  {\bf v}(t + \Delta t) \times {\bf B}({\bf x}(t + \Delta t)) \right].
\end{equation}
The dependence of the magnetic field on the updated positions, ${\bf B}({\bf x}(t + \Delta t))$, is not a problem because we have access to those from the position update in~\refeq{vvdefx}.  Let's focus on isolating ${\bf v}(t+\Delta t)$.  For visual clarity, define ${\bf v} \equiv {\bf v}(t)$, ${\bf B} \equiv {\bf B}({\bf x}(t))$, ${\bf w} \equiv {\bf v}(t+\Delta t)$, ${\bf C} \equiv {\bf B}({\bf x}(t + \Delta t))$, and $\alpha \equiv q \Delta t/(2 m)$, then the vector update from~\refeq{LorentzUp} reads
\begin{equation}\label{wstart}
{\bf w} =\underbrace{ \left( {\bf v} + \alpha {\bf v} \times {\bf B} \right)}_{\equiv {\bf d}} + \alpha {\bf w} \times {\bf C},
\end{equation}
and the term in parenthesis on the right is a constant independent of the target ${\bf w}$; call it ${\bf d}$ as shown.  Dotting ${\bf C}$ into both sides of~\refeq{wstart} gives ${\bf w} \cdot {\bf C} = {\bf d} \cdot {\bf C}$ since ${\bf w} \times {\bf C}$ is perpendicular to ${\bf C}$.  Crossing ${\bf C}$ into both sides of~\refeq{wstart}, we get
\begin{equation}
{\bf w} \times {\bf C}  = {\bf d} \times {\bf C} - \alpha \left( {\bf w} C^2 - {\bf C} ({\bf d} \cdot {\bf C})\right).
\end{equation}
The expression on the right depends linearly on ${\bf w}$.  Using this form for ${\bf w} \times {\bf C}$ back in the original~\refeq{wstart} yields an equation we can use to isolate ${\bf w}$ on one side, with known vector quantities appearing on the other,
\begin{equation}
{\bf w} = \frac{1}{1 + \alpha^2 C^2} \left[ {\bf d} + \alpha {\bf d} \times {\bf C} + \alpha^2 {\bf C} ({\bf d} \cdot {\bf C}) \right].
\end{equation}
The full velocity Verlet update in this specialized setting is
\begin{eqnarray}
{\bf d} &\equiv&  \of{{\bf v}(t) + \frac{q \Delta t}{2 m} {\bf v} (t) \times {\bf B}({\bf x}(t))},  \\
{\bf x}(t + \Delta t) &=&  {\bf x}(t) + {\bf d} \Delta t   \label{mvxup}, \\
{\bf C} &\equiv & {\bf B}({\bf x}(t + \Delta t)), \\
{\bf v}(t + \Delta t) &=& \frac{1}{1 + \of{\frac{q \Delta t}{2 m}}^2 {\bf C} \cdot {\bf C} } \left[{\bf d} + \frac{q \Delta t}{2 m} {\bf d} \times {\bf C} +   \of{\frac{q \Delta t}{2 m} }^2 {\bf C} \of{ {\bf d} \cdot {\bf C}} \right] \label{mvvup}.
\end{eqnarray}
If we are given initial values, ${\bf x}(0) = {\bf x}_0$ and ${\bf v}(0) = {\bf v}_0$, we can use~\refeq{mvxup} and~\refeq{mvvup} to generate approximations to ${\bf x}(\Delta t)$ and ${\bf v}(\Delta t)$, then use those values to obtain ${\bf x}(2 \Delta t)$ and ${\bf v}(2 \Delta t)$, and so on, up to any desired final time.

The pseudocode for the magnetic velocity Verlet method is shown in Algorithm 1 below.  You provide the function ``MVVerlet" with: 1.\ the initial position vector, ${\bf x}_0$, 2.\ the initial velocity vector, ${\bf v}_0$, 3.\ the particle mass $m$, 4.\ the particle charge $q$, 5.\ the time step size $\Delta t$, 6.\ the number of steps to take, $n$, and 7.\  a function ${\bf B}({\bf x})$ that returns the magnetic field at the point ${\bf x}$.  The method returns a list of the particle's position ($X$) and velocity ($V$) where the list index $j$ is associated with time $(j-1) \Delta t$ for $j = 1 \rightarrow n$.
\begin{algorithm}[H]
\setstretch{1.}
\caption{MVVerlet$({\bf x}_0, {\bf v}_0, m, q, \Delta t, n, {\bf B})$}
\label{alg:test}
\begin{algorithmic}
\STATE $X \gets$ length $n$ list of zeroes
\STATE $X_1 \gets {\bf x}_0$ 
\STATE $V \gets$ length $n$ list of zeroes
\STATE $V_1 \gets {\bf v}_0$
\STATE $\alpha \gets q \Delta t/(2 m)$
\STATE ${\bf x}  \gets {\bf x}_0$
\STATE ${\bf v} \gets {\bf v}_0$
\FOR{$j=2 \rightarrow n$}
\STATE $\meskip {\bf d} \gets {\bf v} + \alpha {\bf v} \times {\bf B}({\bf x})$
\STATE $\meskip {\bf x} \gets {\bf x} +  {\bf d} \Delta t$
\STATE $\meskip {\bf C} \gets {\bf B}({\bf x})$
\STATE $\meskip {\bf v} \gets ({\bf d} + \alpha {\bf d} \times {\bf C} + \alpha^2 {\bf C} ({\bf d} \cdot {\bf C}) )/(1 + \alpha^2 C^2) $
\STATE $\meskip X_j \gets {\bf x}$
\STATE $\meskip V_j \gets {\bf v}$
\ENDFOR
\RETURN{$\{X,V\}$}
\end{algorithmic}
\end{algorithm}

\section{Uniform Circular Motion}

\begin{figure}[htbp] 
   \centering
   \includegraphics[width=3in]{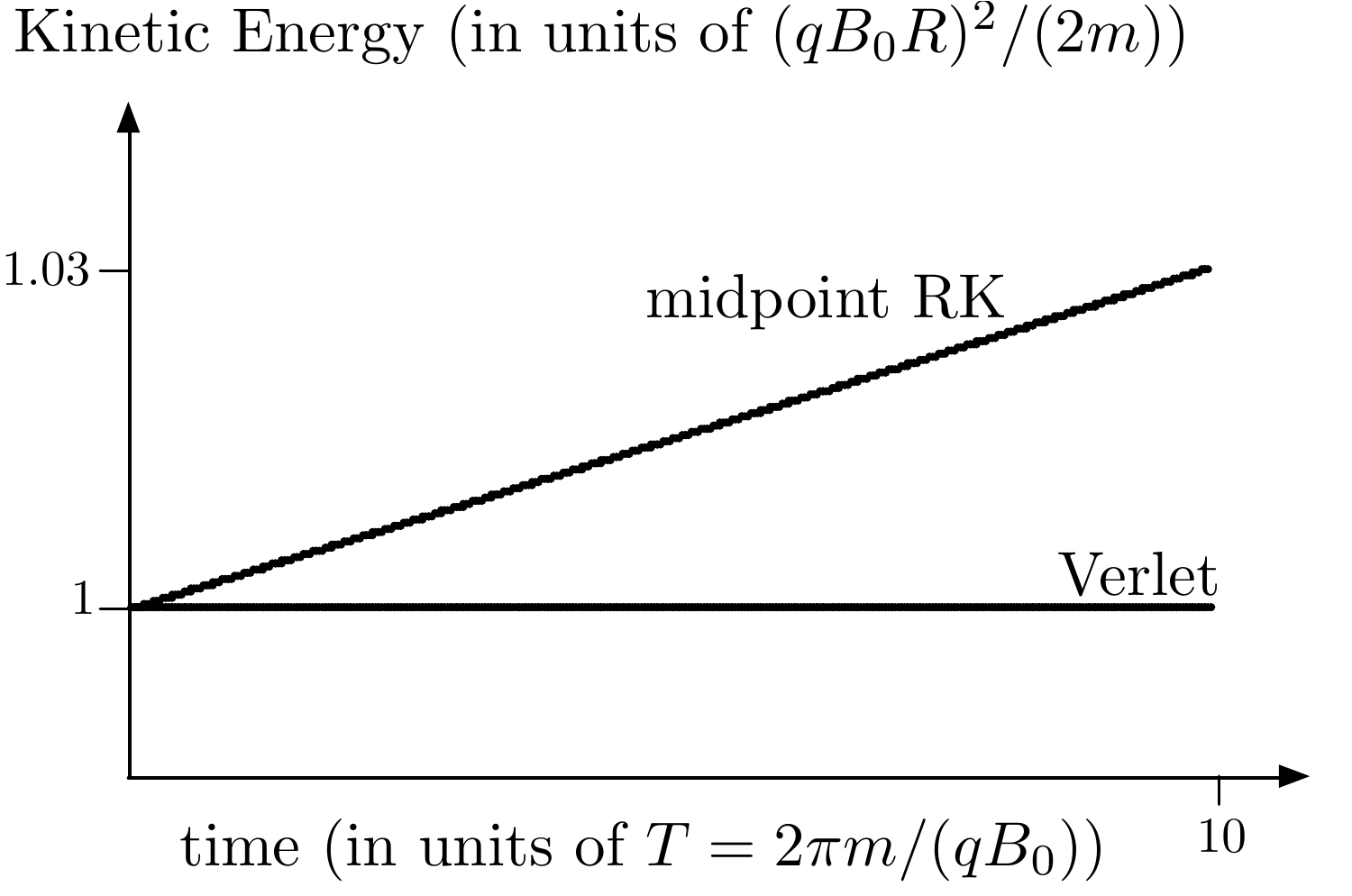} 
   \caption{The kinetic energy as a function of time for a charged particle moving in a uniform magnetic field as calculated by the magnetic Verlet method and the midpoint Runge-Kutta method.}
   \label{fig:KEcomp}
\end{figure}

To test the method and display its energy conservation, take a uniform magnetic field ${\bf B} = B_0 \hat{\bf z}$.  A particle of mass $m$ and charge $q > 0$ starts at initial position ${\bf x}_0 = R \hat{\bf x}$ with initial velocity ${\bf v}_0 = -q B_0 R/m \hat{\bf y}$ and moves in a circle with period $T = 2 \pi m/(q B_0)$.  We'll approximate the trajectory for ten cycles, with a time step of $\Delta t = T/50$ using both the magnetic Verlet approach from above and a midpoint Runge-Kutta method with the same initial conditions and time step.  A plot of the kinetic energy for each method as a function of time is shown in~\reffig{fig:KEcomp}.  The difference between the maximum kinetic energy and minimum kinetic energy, divided by the initial kinetic energy, is $\sim 10^{-14}$ for the Verlet method, compared with $10^{-2}$ for midpoint Runge-Kutta.

The uniform circular motion here defines a constant magnetic dipole moment.  In general, the dipole moment for a current density ${\bf J}(\vecbf x)$ is defined by the volume integral over all space,
\begin{equation}
{\bf m} = \frac{1}{2} \int {\bf x} \times {\bf J(\vecbf x)} d\tau.
\end{equation}
For a particle at location ${\bf r}(t)$ at time $t$, the current density is ${\bf J}({\bf x}) = q \delta^3({\bf x} - {\bf r}(t)) \dot{\bf r}(t)$, and the dipole moment becomes ${\bf m} = q {\bf r}(t) \times {\bf v}(t)/2$. 
In cylindrical coordinates, $\{ s, \phi, z\}$, with motion in the $xy$ plane and ${\bf r} \perp {\bf v}$ as it is here, the moment is ${\bf m} = -(q v_\phi s/2) \hat{\bf z}$ where the minus sign comes from the direction of circulation, $\vecbf v = -v_\phi \hat{\bm \phi}$ (clockwise motion).  The uniform circular trajectory has $s = R$ and $v_\phi = q B_0 R/m$ so the particle's dipole moment is 
\begin{equation}\label{ucmdip}
{\bf m} =-\frac{(q R)^2 B_0}{2 m} \hat{\bf z}.
\end{equation}

\begin{figure}[h] 
   \centering
   \includegraphics[width=2.75in]{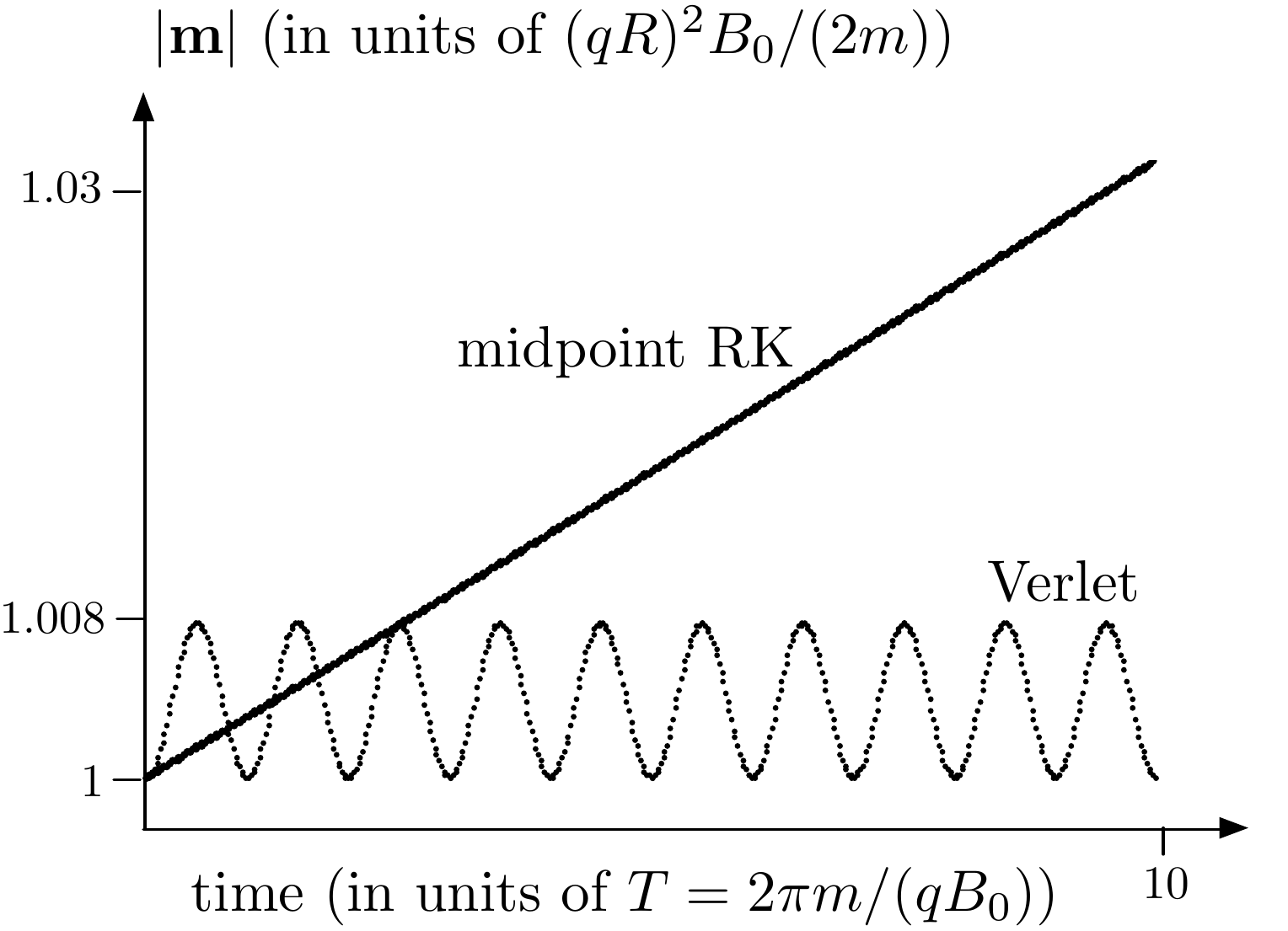} 
   \caption{The magnitude of the magnetic dipole moment of the particle undergoing uniform circular motion as a function of time.  The midpoint Runge-Kutta method shows linear growth in $|{\bf m}|$, while the Verlet method has much smaller, bounded oscillatory error with period that is the same as the period of the circular motion.}
   \label{fig:moment}
\end{figure}

The magnitude of the magnetic dipole moment is constant, but how well do the numerical methods preserve its value?  Working in units of $(q R)^2 B_0/(2 m)$, the magnitude of the magnetic moment as a function of time is calculated for both the magnetic Verlet method and the midpoint Runge-Kutta method.  The results are shown in~\reffig{fig:moment}, where it is clear that the Verlet method has dipole magnitude with bounded error as time goes on, while the midpoint Runge-Kutta method has error that grows linearly with time, similar to its treatment of the kinetic energy.  Overall, the magnetic velocity Verlet method is superior in preserving these constants of the motion for this simplest test case. 

The dipole moment magnitude can be written in terms of the particle's kinetic energy in the direction perpendicular to the magnetic field.  For the uniform circular motion in this example, the perpendicular kinetic energy is
\begin{equation}
K_\perp = \frac{1}{2} m \of{\frac{q B_0 R}{m}}^2.
\end{equation}
Comparing this expression with the magnitude of the dipole moment in~\refeq{ucmdip}, it is the combination $K_\perp/B_0$ that captures $|{\bf m}|$.

In settings where the longitudinal magnetic field does not change magnitude much over some region, particle trajectories follow roughly circular motion in the perpendicular plane.  Then
\begin{equation}\label{mudef}
\mu \equiv \frac{K_\perp}{B},
\end{equation}
which is constant for uniform $B_0$, is approximately constant, and is called the ``first adiabatic invariant."~\cite{JACKSON,VANALLEN,GOLDSTEIN} 
An equivalent expression for this constant, motivated by uniform circular motion where it follows from~\refeq{mudef}, is
\begin{equation}\label{mualt}
\mu = \frac{1}{2} q v_\perp s
\end{equation}
for $s$, the radius of the circular motion with speed $v_\perp$.
 We can use the approximately constant value of $\mu$ to develop the mirror motion of a particle moving through a non-uniform longitudinal field.

\section{Cylindrical Magnetic Mirror Machine}

Take a longitudinal magnetic field component pointing in the $z$ direction with magnitude that changes as a function of $z$, $B_z(z)$.  As a charged particle moves in the $z$ direction, it is presented with a series of field magnitudes that are uniform in the $xy$ plane, and will undergo roughly circular motion in that plane.  To get an approximately constant value of $\mu$, the longitudinal motion must be slow enough that multiple cycles of the approximately circular motion occur before the longitudinal magnetic field changes significantly.  This is the ``adiabatic assumption," that the particle's longitudinal motion is slower than its perpendicular motion.  We will check that this is satisfied, and also that $\mu$ is constant, in the numerical solutions below.

A magnetic field cannot consist of $B_z(z) \hat{\bf z}$ alone; that violates $\nabla \cdot {\bf B} = 0$.  In order to preserve azimuthal symmetry, introduce a magnetic field component pointing in the $s$ direction that depends on $s$ and $z$, ${\bf B} = B_z(z) \hat{\bf z} + B_s(s,z) \hat{\bf s}$.  To make the field divergenceless, $B_s(s,z)$ must be related to $B_z(z)$ by
\begin{equation}\label{Bsdiv}
B_s(s,z) = -\frac{s}{2} \frac{dB_z(z)}{d z}.
\end{equation}
The longitudinal force on the particle is no longer zero.  For the clockwise circulation associated with a positive charge, $\vecbf v_\perp = -v_\phi \hat{\bm \phi}$, and the equation of motion in the $z$ direction is
\begin{equation}\label{zeom}
m \ddot z(t) = q v_\phi B_s = -\frac{q v_\phi s}{2}  \frac{d B_z}{d z}.
\end{equation}
The term sitting out front in the second equality is precisely $\mu$ from~\refeq{mualt} with $v_\perp = v_\phi$, so that 
\begin{equation}
m \ddot z(t) = -\mu \frac{d B_z}{d z}.  
\end{equation}
Multiplying both sides of this equation by $\dot z(t)$, 
\begin{equation}
m \dot z(t) \ddot z(t) =-\mu  \frac{d B_z}{d z} \frac{d z(t)}{d t},
\end{equation}
we can write both sides as total time derivatives provided $\mu$ is constant, 
\begin{equation}
\frac{d}{dt} \of{ \frac{1}{2} m \dot z(t)^2 } = -\frac{d}{dt} \of{\mu B_z(z(t))}.
\end{equation}
The integration is easy to carry out, giving
\begin{equation}\label{Kcons}
\frac{1}{2} m \dot z(t)^2 + \mu B_z(z(t) ) = C
\end{equation}
where $C$ is a constant of integration to be set by the initial conditions.

If a particle starts off at $z(0) = z_0$, with $\dot z(0) = 0$,  then $C = \mu B_z (z_0)$, and~\refeq{Kcons} becomes
\begin{equation}\label{longKE}
\frac{1}{2} m \dot z(t)^2 = -\mu \of{ B_z(z(t)) - B_z(z_0)}.
\end{equation}
In order for the longitudinal speed to be real, $B_z(z(t))$ must be less than the initial value of $B_z(z_0)$.  The particle will move towards regions of smaller field, increasing its longitudinal speed while circulating in the $xy$ plane.  If the longitudinal magnetic field magnitude increases towards the value of $B_z(z_0)$ at some location, the particle's longitudinal speed will decrease.  For a location $z_1$ with $B_z(z_1) = B_z(z_0)$, the particle must have zero longitudinal speed.  The longitudinal kinetic energy in~\refeq{longKE} plays a role similar to a one dimensional energy conservation equation in classical mechanics.  We can use it to predict some features of motion in the longitudinal direction, like turning points and points of maximum speed.

Using~\refeq{longKE}, it is easy to see how to construct $B_z(z)$ so as to get oscillatory longitudinal motion:  Make a field that is symmetric about $z = 0$ with maxima at $\pm z_0$.  Then a particle that starts with no longitudinal speed at $z_0$ will also have zero longitudinal velocity component at $-z_0$.  So the motion of the particle will be confined between these two points.  This ``mirror machine" configuration was originally introduced as a way to confine plasmas without having mechanical pieces in contact with the plasma.\cite{POST}

\begin{figure}[htbp] 
   \centering
   \includegraphics[width=3in]{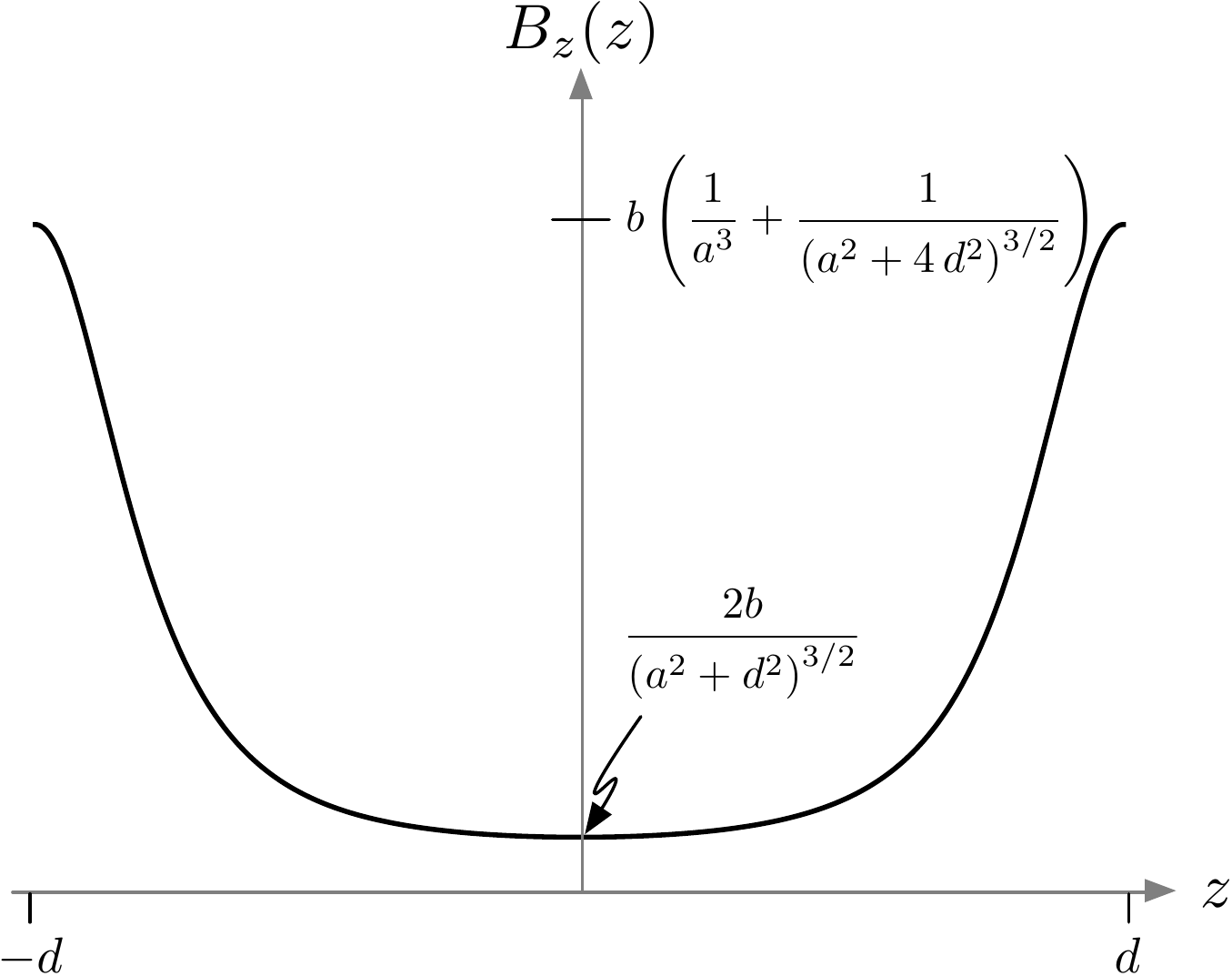} 
   \caption{The magnitude of the magnetic field from~\refeq{Bzonly}.  The current rings are at locations $\pm d$, and the maximum and minimum field values depend on both $d$ and the radius of the rings, $a$, in addition to the size of the steady current, enapsulated in $b$.}
   \label{fig:Bzplot}
\end{figure}

A simple, physically inspired way to achieve a magnetic mirror machine field is to put a pair of current loops of radius $a$ carrying steady current $I$ at locations $\pm d$ along the $z$ axis.  The field produced by this configuration, at $z$ along the axis is
\begin{equation}\label{Bzonly}
{\bf B} = \underbrace{\frac{\mu_0 I a^2}{2}}_{\equiv b} \hat{\bf z} \left[ \frac{1}{ \of{ (d - z)^2 + a^2}^{3/2} } + \frac{1}{ \of{ (d + z)^2 + a^2}^{3/2} } \right],
\end{equation}
with field magnitude shown in~\reffig{fig:Bzplot}.

For motion occurring near the $z$ axis, this field is approximately valid, and we'll take it to define the longitudinal component of the target mirror field. The requirement in~\refeq{Bsdiv} then provides the radial component.  Thus, the idealized magnetic field is~\cite{INTNOTE}
\begin{equation}\label{mirrorfield}
\begin{aligned}
{\bf B}  &=  b \left[ \frac{1}{ \of{ (d - z)^2 + a^2}^{3/2} } + \frac{1}{ \of{ (d + z)^2 + a^2}^{3/2} } \right]  \hat{\bf z} \\
&- \frac{3 b s}{2} \left[ \frac{d-z}{\of{a^2 + (d-z)^2}^{5/2}} - \frac{d+z}{\of{a^2 + (d+z)^2}^{5/2} }\right] \hat{\bf s}.
\end{aligned}
\end{equation}

\begin{figure}[htbp] 
   \centering
\subfigure[\, The particle's $z$ position as a fraction of the initial $z(0) = d/4$.]{ \includegraphics[width=2.5in]{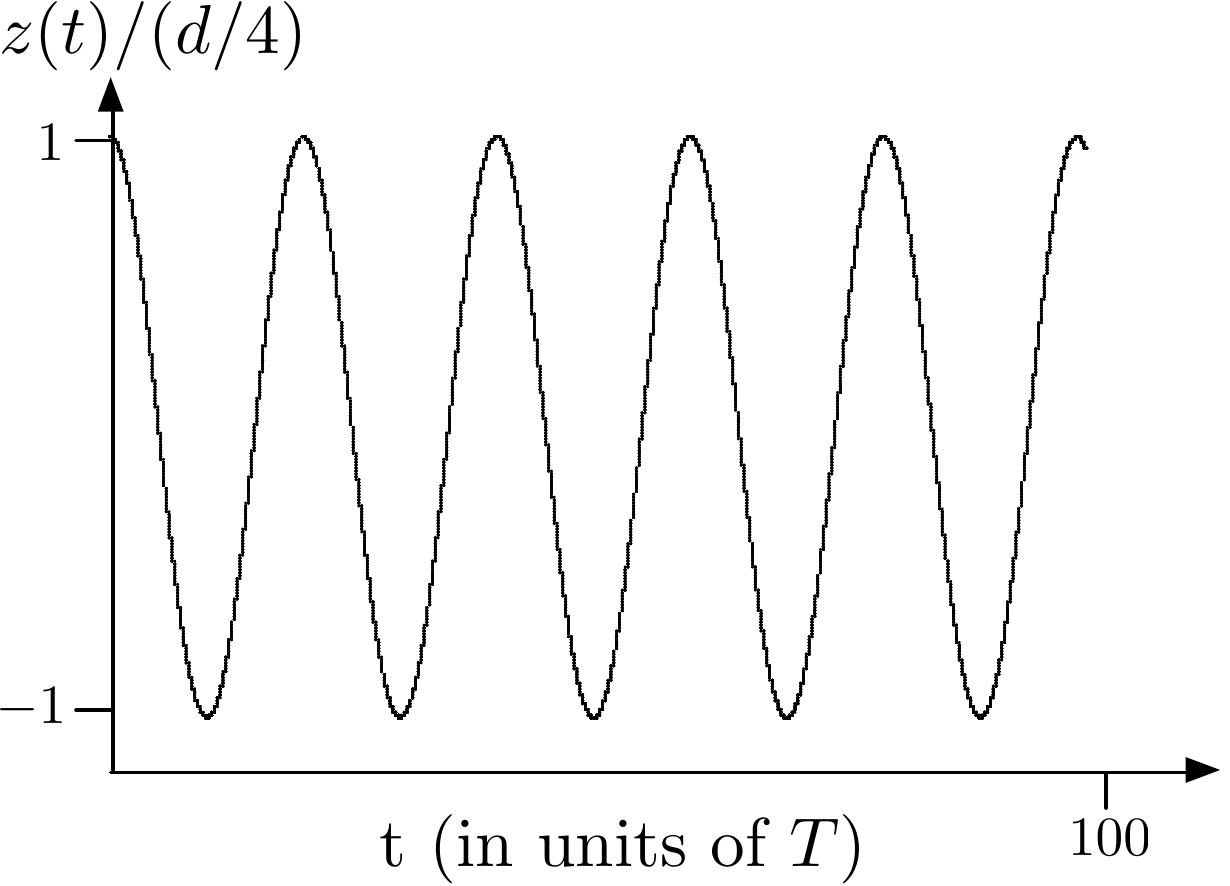} } \qquad
\subfigure[\, The particle's radial distance to the $z$ axis as a fraction of the initial radius $R$.]{ \includegraphics[width=2.5in]{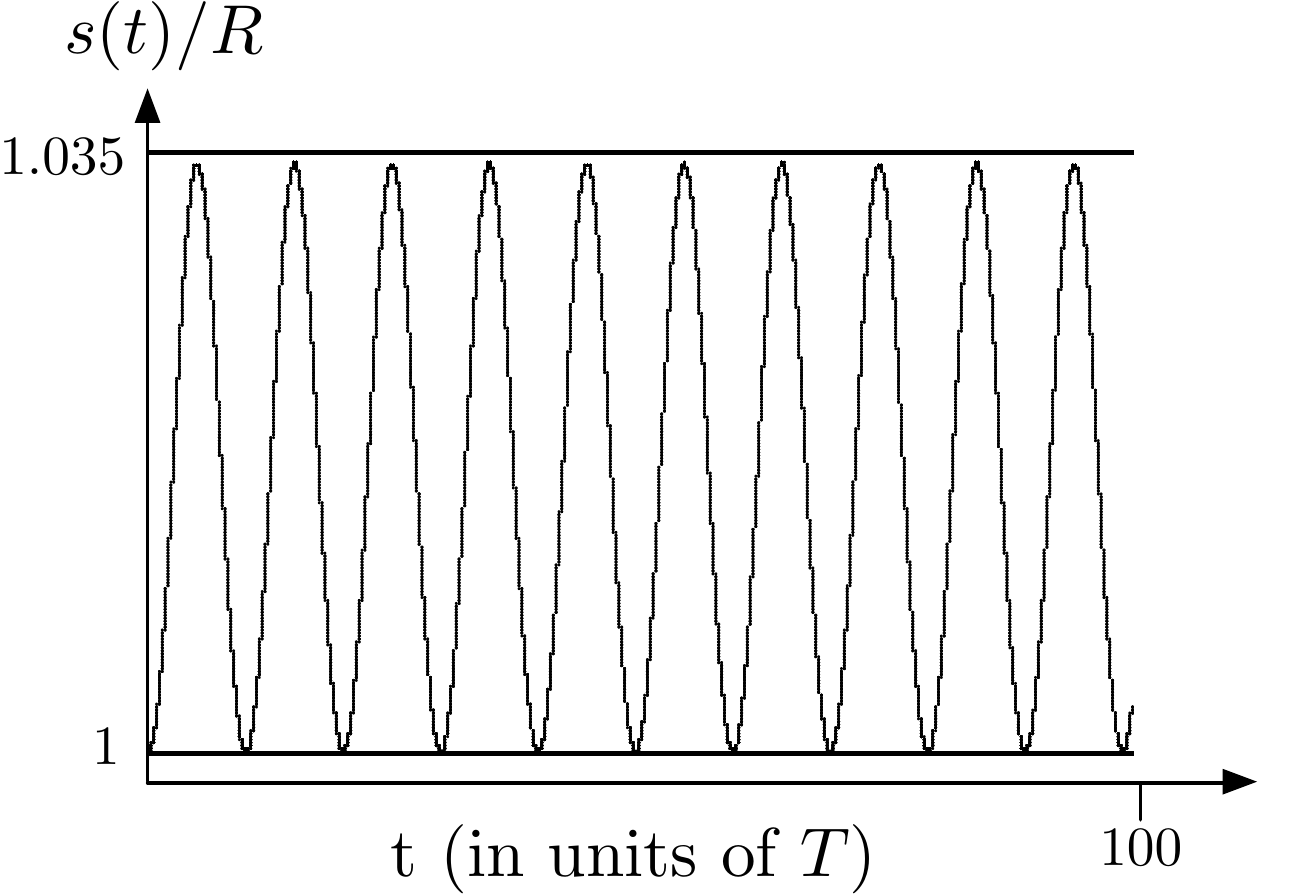}} \\
\subfigure[\, The particle trajectory.]{ \includegraphics[width=1.in]{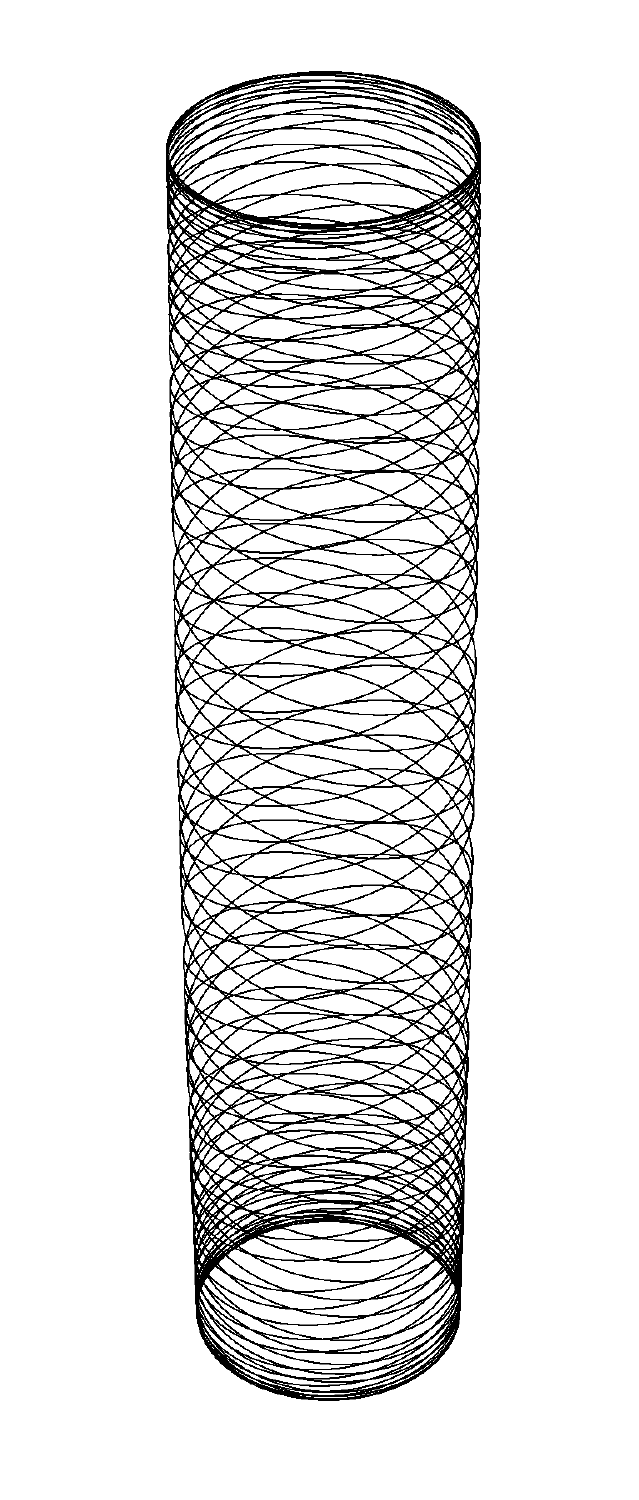} }  
\caption{Properties of the numerical solution using the initial conditions in~\refeq{poszero} and~\refeq{velzero}.}
   \label{fig:d4data}
\end{figure}

Starting a particle off at $z(0) = z_0$ with $\dot z(0) = 0$, we expect it to move back and forth between $\pm z_0$.
We can take a look at one such trajectory to see how well the kinetic energy and the value of $\mu$ are conserved numerically.  For the  initial position and velocity, take
\begin{eqnarray} 
{\bf r}(0) &=& R \hat{\bf x} + \frac{d}{4} \hat{\bf z} \label{poszero} \\
{\bf v}(0) &=& -\frac{q B_z(d/4) R}{m} \hat{\bf y}, \label{velzero}
\end{eqnarray}
where $R = a/20$ is chosen to be small compared to the radius of the current loops.  With these initial conditions, if the longitudinal field were uniform with constant value $B_z(d/4)$, the particle would undergo circular motion of radius $R$ with period
\begin{equation}
T = 2 \pi \frac{m}{q B_z(d/4)}.
\end{equation}
 Running the magnetic Verlet method for a total time of $100 T$ in steps of $\Delta t = T/200$, it is clear from the position plot in~\reffig{fig:d4data} that the motion in $z$ is periodic, going back and forth between $\pm d/4$.  That period of oscillation is much larger than the period of the perpendicular circular motion,  with roughly five full cycles along the $z$ axis occurring over the $100T$ time frame.  The longitudinal motion is slower than the circular motion, so the adiabatic assumption is satisfied here.  Another piece of our assumption was that the radius of the circular motion does not change much over the course of the trajectory, and that is true here, as shown in Fig. 4(b).

\begin{figure}[htbp] 
   \centering
   \includegraphics[width=3.5in]{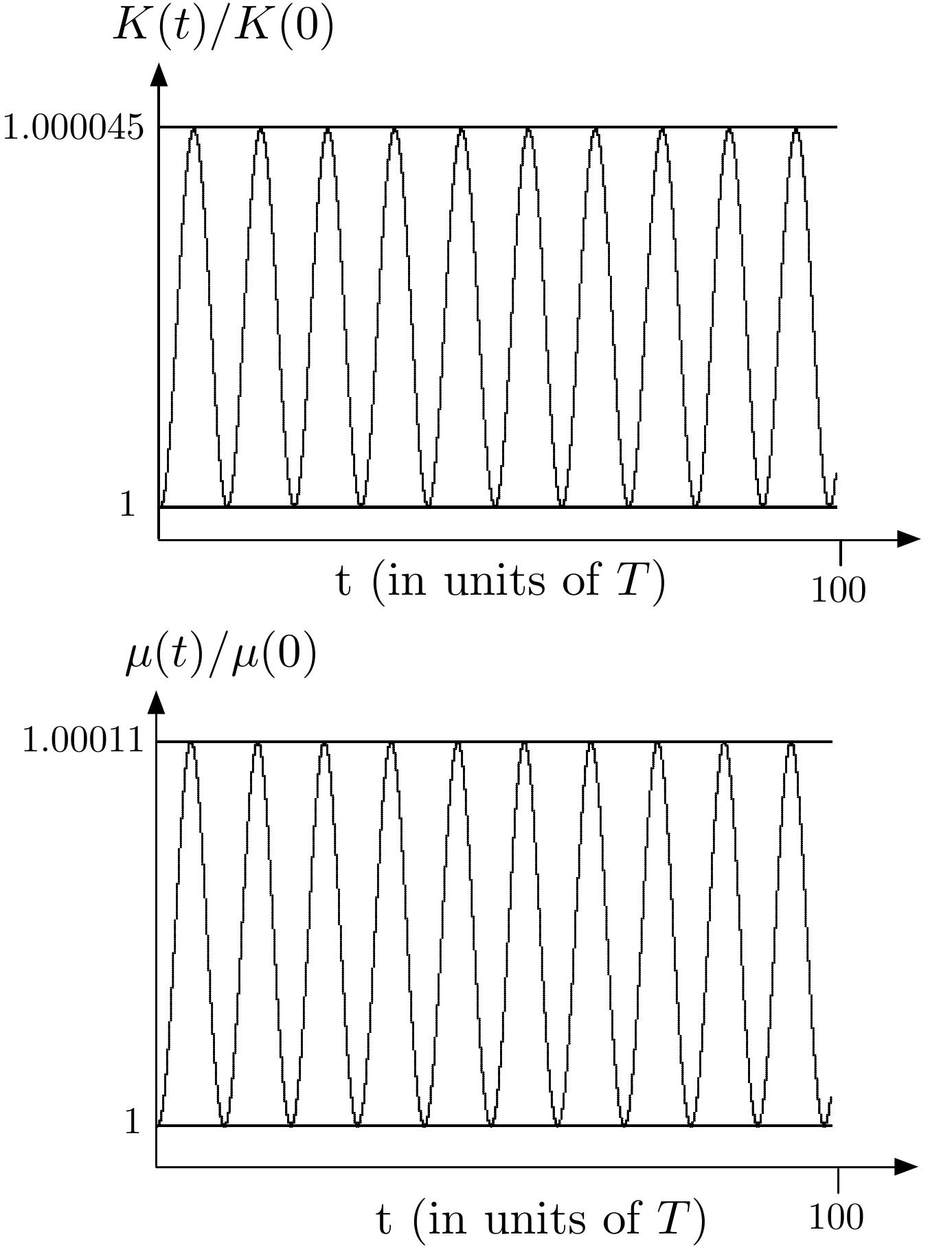} 
   \caption{The kinetic energy (top) and adiabatic constant $\mu$ (bottom), both displayed as fractions of their initial values.  These show that the kinetic energy is quite well conserved, and the value of $\mu$ approximately conserved as a particle oscillates between $-d/4$ and $d/4$ along the $z$ axis while executing (roughly) circular motion in the $xy$ plane.}
   \label{fig:d4em3}
\end{figure}

 In~\reffig{fig:d4em3}, we've plotted the kinetic energy and value of $\mu$ from~\refeq{mudef} using $B_z \approx B$ for the denominator since $B_s$ is small.  In both of these plots, it is the ratio with the initial value that gives us a dimensionless measure of the numerical constancy. The kinetic energy, which is strictly conserved by the equations of motion, sets the numerical standard for a ``constant of the motion" with $\sim .005\%$ change over the timescale shown.  The adiabatic constant $\mu$ is approximately conserved exhibiting larger $\sim .01\%$ change over the $100 T$ time frame.

\begin{figure}[htbp] 
   \centering
   \subfigure[\, The particle trajectory]{\includegraphics[width=2in]{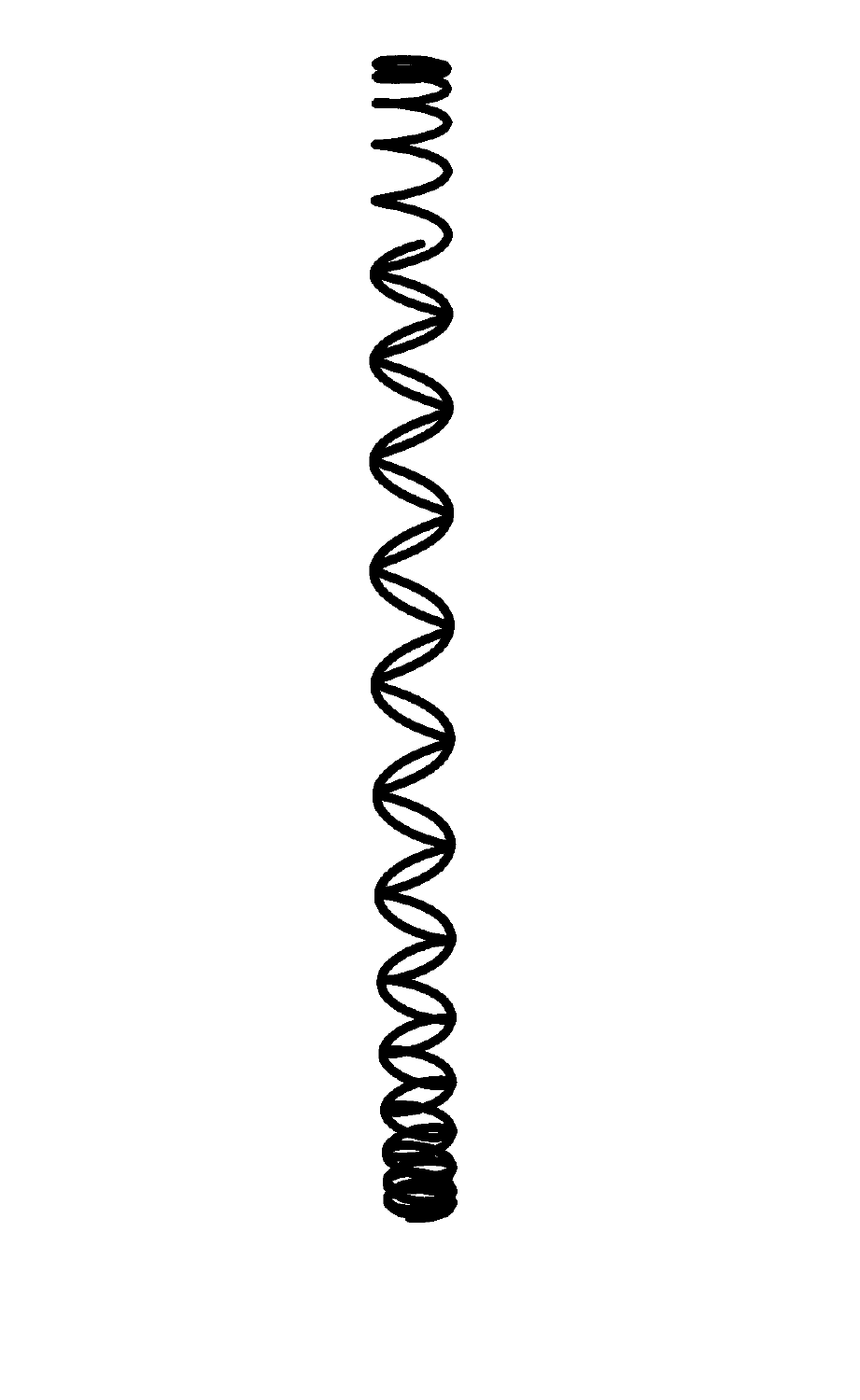} }
\subfigure[\, The kinetic energy (top) and adiabatic constant (bottom) as fractions of their initial values.]{\includegraphics[width=2.5in]{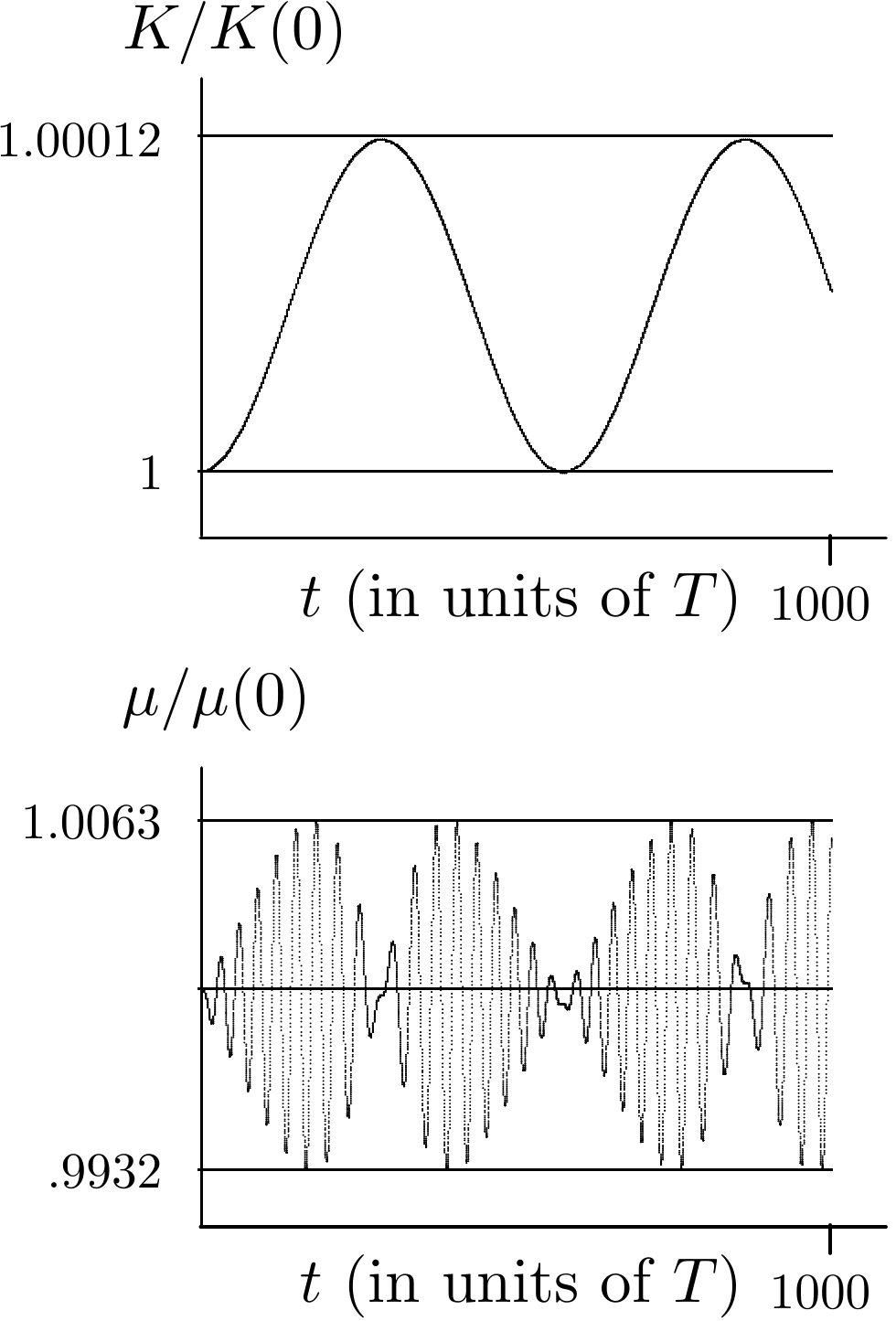}}
 \caption{\, Numerical trajectory information for a particle with $v_0 = \alpha$.}
    \label{fig:trajes2}
\end{figure}

\begin{figure}[htbp] 
   \centering
   \subfigure[\, The particle trajectory.]{\includegraphics[width=2in]{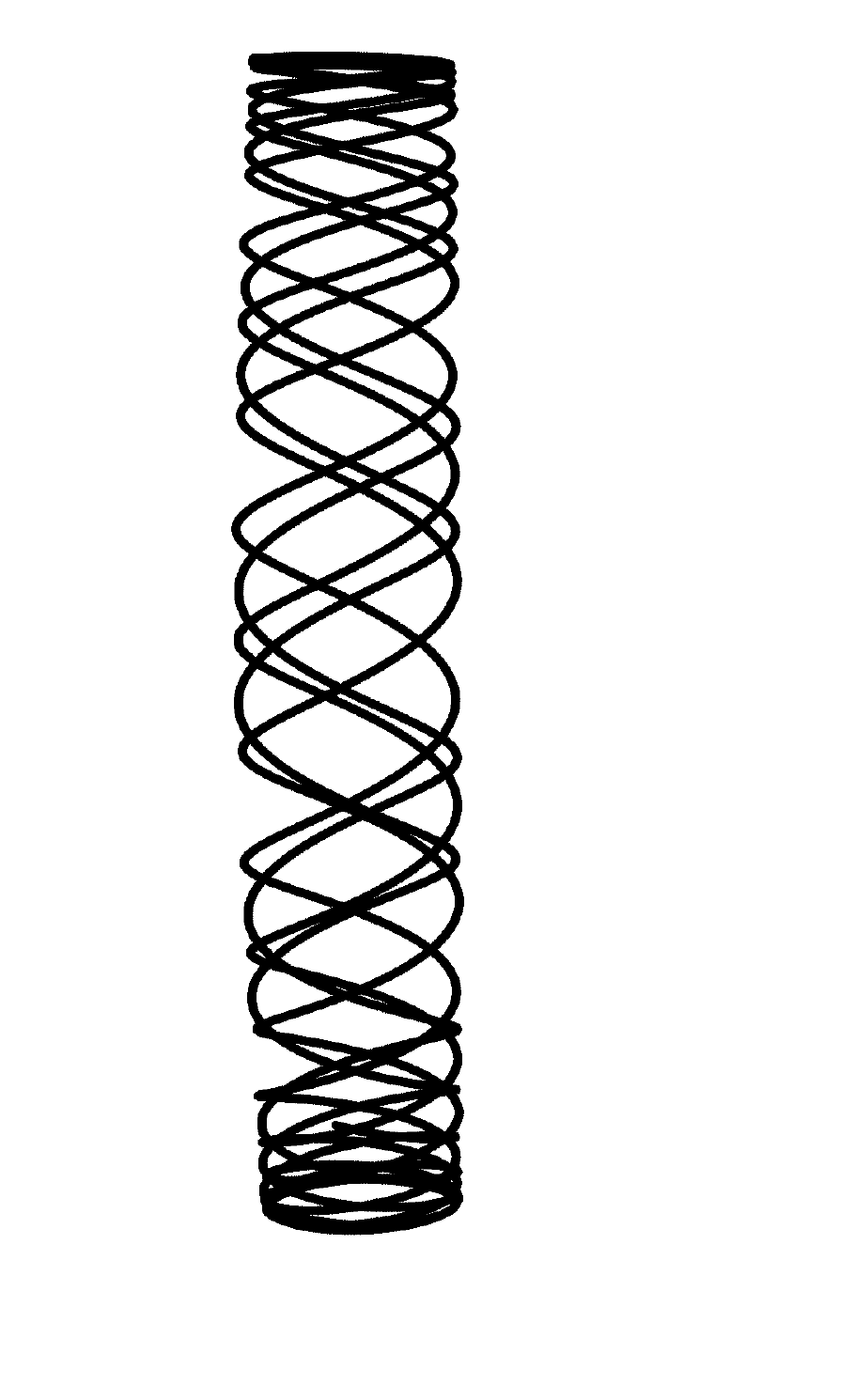} }
\subfigure[\, The kinetic energy (top) and adiabatic constant (bottom) as fractions of their initial values.]{\includegraphics[width=2.5in]{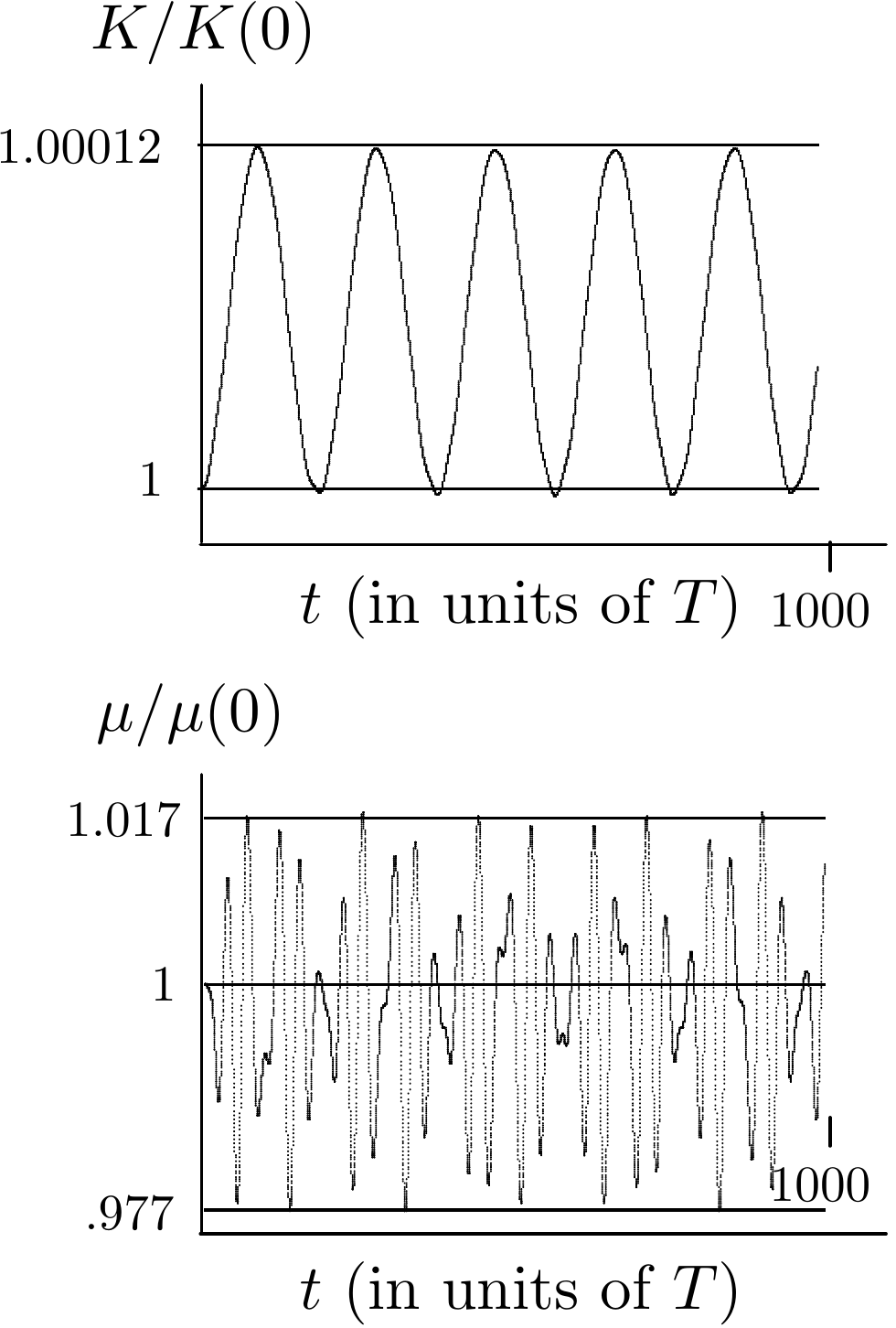}}
 \caption{Numerical trajectory information for a particle with $v_0 = 3 \alpha$.}
    \label{fig:trajes3}
\end{figure}

The initial velocity in~\refeq{velzero} used a very specific tuning for its perpendicular component.  That velocity was taken to be the value that would ensure uniform circular motion of radius $R$ in a uniform magnetic field.  But changing the initial value of $v_\perp$ just changes the radius of the circular motion.~\cite{PROVIDED}  Starting from an initial position ${\bf r}(0) = (d/2) \hat{\bf z}$ with velocity ${\bf v}(0) = -v_0 \hat{\bf x}$, we can define the time-scale
\begin{eqnarray}
R &\equiv& \frac{m v_0}{q B_z(d/2)}, \\
T &\equiv& \frac{2 \pi R}{v_0}.
\end{eqnarray}
Then using a time step of $\Delta t = T/200$, we made two trajectories with different values of $v_0 = 2 \pi R/T$.  The first, with $v_0 =\alpha \equiv 2 \pi R_0/T$ for $R_0 \approx .06$ m, has properties shown in~\reffig{fig:trajes2}, and the second, with a value of $v_0 = 3 \alpha$, has properties shown in~\reffig{fig:trajes3}.  The larger speed leads to a larger radius for the circular motion of the particle shown in~\reffig{fig:trajes3}.  In both cases, the kinetic energy is well conserved, while the first adiabatic invariant shows better conservation in the first case where the motion is closer to the $z$ axis.

\section{Dipole Magnetic Mirror}
\begin{figure}[htbp] 
   \centering
   \includegraphics[width=4in]{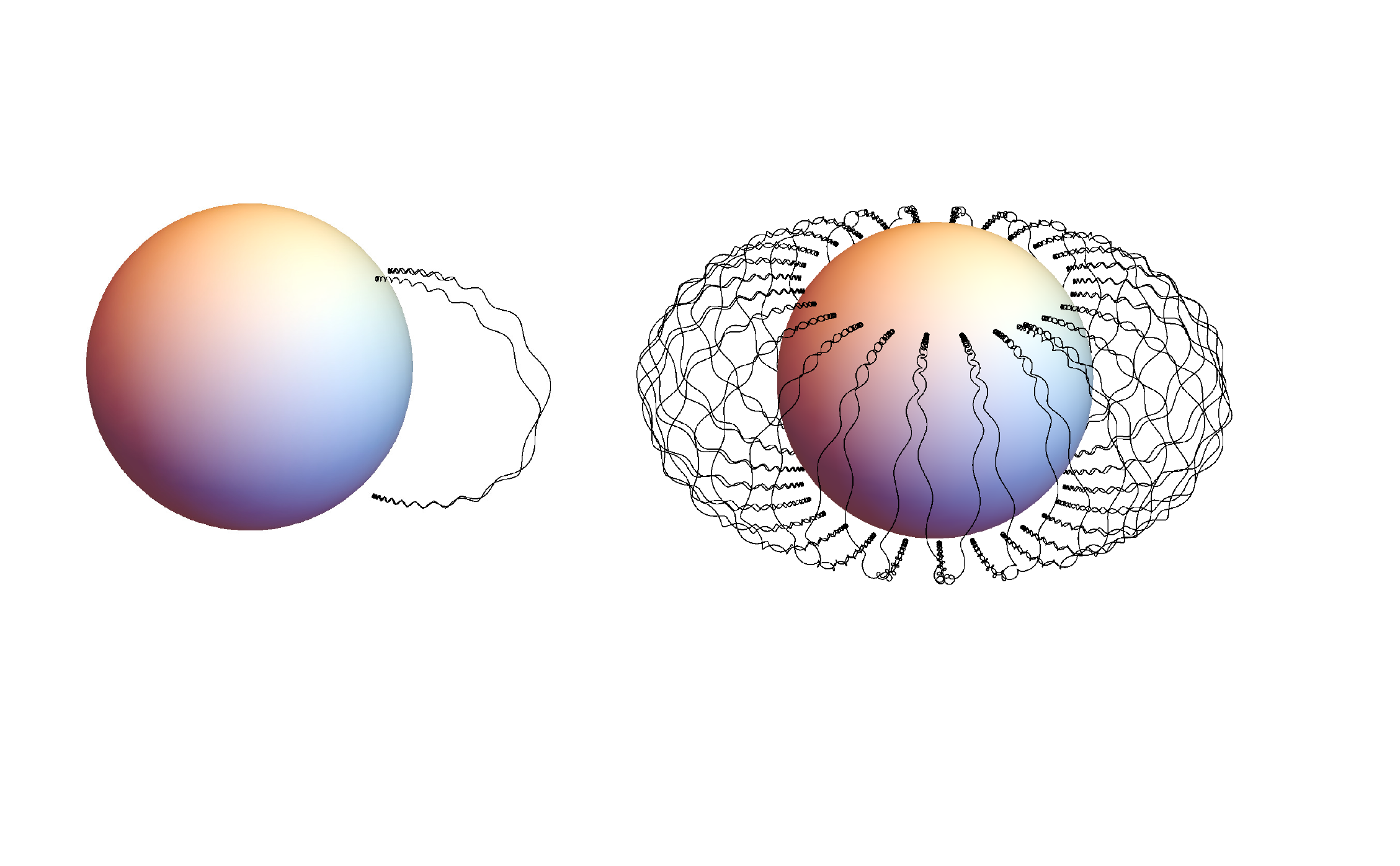} 
   \caption{The mirror effect for a dipolar field.  On the left we see a cycle of the particle moving from the northern to southern hemisphere and back.  On the right, many cycles of the same motion.  The sphere is shown just for scale; it has nothing to do with the dipolar field source.}
   \label{fig:dipole}
\end{figure}

The spiraling, oscillatory motion we saw in the cylindrical setting persists in more complicated field geometries.  Imagine bending a magnetic field line from the previous section so that it curves.  We can still get particle motion that follows the curving field line, circling around it, while encountering regions of increasing magnetic field that act as mirrors.  Consider, for example, a dipolar magnetic field like the one outside the earth.  For a magnetic dipole pointing in the $z$ direction, the field is 
\begin{equation}
{\bf B} = \frac{k}{4 \pi \of{x^2 + y^2 + z^2}^{5/2}} \left[ 3 x z \hat{\bf x} + 3 y z \hat{\bf y} - (x^2 + y^2 - 2 z^2) \hat{\bf z} \right],
\end{equation}
where the constant $k$ sets the magnitude.  We ran the magnetic Verlet method for a particle with initial position in the $yz$ plane using a polar angle of $\theta_0 = 50^\circ$, with initial velocity in the $x$ direction.  The result is shown in~\reffig{fig:dipole}, where we can see one ``cycle" of the mirrored trajectory, and then multiple cycles, going all the way around.

It is interesting that the particle trajectory follows the field lines, spiraling tightly around them as it goes up and back.  This trajectory allows us to map out the dipole field visually using the motion of the charged particle.  For electric forcing, field lines point in the direction of acceleration of the particle, and if we start a particle from rest, charges travel along these lines.  We build intuition about electric fields using this property.  Magnetic field lines are perpendicular to the velocity vector, which points in the direction of motion, so we are not used to mapping magnetic field lines using the trajectories they generate.  The type of motion shown in~\reffig{fig:dipole}, induced by the magnetic field of the earth, is what causes the Van Allen belts of particles trapped by the earth's field, traveling up and back along its field lines.~\cite{VANALLEN,MAPLES}
 
\section{Monopole Field}

As our final example, we study the motion of a charged particle in the presence of a magnetic monopole field (or, if you prefer, very close to the north pole of a dipole field).  We will again see the behavior that has been the focus of this paper:  a charge follows a field line while corkscrewing around it.  As the particle encounters a region of increasing field magnitude, its motion along the field line slows, eventually stopping and reversing direction along the line.  This time, since the field lines converge radially, the geometry of the circulation perpendicular to a field line is conical.  The magnetic monopole case has the advantage that there is a closed form expression for the motion of the charge (see Griffiths' Problem 5.45 and references there~\cite{GRIFFITHS}), giving us a rare exact result with which to compare the numerical trajectory.

For a magnetic monopole, the field is ${\bf B} = k \hat{\bf r}/(4 \pi r^2)$ where $k \equiv \mu_0 q_m$ is a constant that is set by the charge of the monopole.  As usual, the kinetic energy of a charged particle moving under the influence of this field is constant.  In addition, there is a conserved vector
\begin{equation}
{\bf Q} \equiv m \left( {\bf r} \times {\bf v} \right) - \frac{k q}{4\pi} \hat{\bf r},
\end{equation}
and this vector can be aligned so that it points along the $z$ axis: ${\bf Q} = Q_0 \hat{\bf z}$.  Using the constancy of ${\bf Q}$ and the kinetic energy, one can show that $\theta(t) \equiv \theta$ is a constant of the motion, and develop expressions for the time derivatives of $r(t)$ and $\phi(t)$:
\begin{eqnarray}
\dot r(t) &= & \pm \sqrt{ v^2 - \left(\frac{Q_0 \sin\theta}{m r(t)}\right)^2 },  \label{dotr} \\
 \dot\phi(t) &= & \frac{Q_0}{m r(t)^2}, \label{dotphi}
\end{eqnarray}
where $v$ is the constant speed of the particle.  Dividing $\dot r(t)$ by $\dot \phi(t)$, the ODE governing the spherical $r$ coordinate parametrized by $\phi$ is
\begin{equation}
\frac{d r(\phi)}{d \phi} =\pm \sqrt{v^2 - \left(\frac{Q_0 \sin\theta}{m r(\phi)} \right)^2} \left(\frac{m r(\phi)^2}{Q_0}\right).
\end{equation}
This equation can be solved, taking the minus sign, 
\begin{equation}\label{rofphi}
r(\phi) =- \frac{Q_0}{m v} \frac{\sin\theta}{\cos \left( (\phi - \alpha) \sin\theta\right)},
\end{equation}
where $\alpha$ is the constant of integration.  We have lost the temporal evolution, but this equation can be used to draw a picture of the trajectory, and we can compare that with the solution to the equations of motion that we get numerically from the magnetic Verlet method.  The numerical solution requires a complete set of initial conditions.  We can get those by first choosing the constants of motion, $\theta$ (we took $\theta = 4^\circ$), $\alpha = -8$ (chosen to give a trajectory that moved radially inward initially), and $Q_0$ (negative one, related to the size of the magnetic monopole).  For the initial value of $\phi$, it is convenient to start at $\phi = 0$, and then the initial value of $r(\phi = 0)$ is given by the solution in~\refeq{rofphi}.  Finally, the initial values for $\dot r(t = 0)$ and $\dot \phi(t=0)$ can be obtained from~\refeq{dotr} and~\refeq{dotphi} respectively, using the rest of the initial coordinate values and constants.

\begin{figure}[htbp] 
   \centering
   \includegraphics[width=3in]{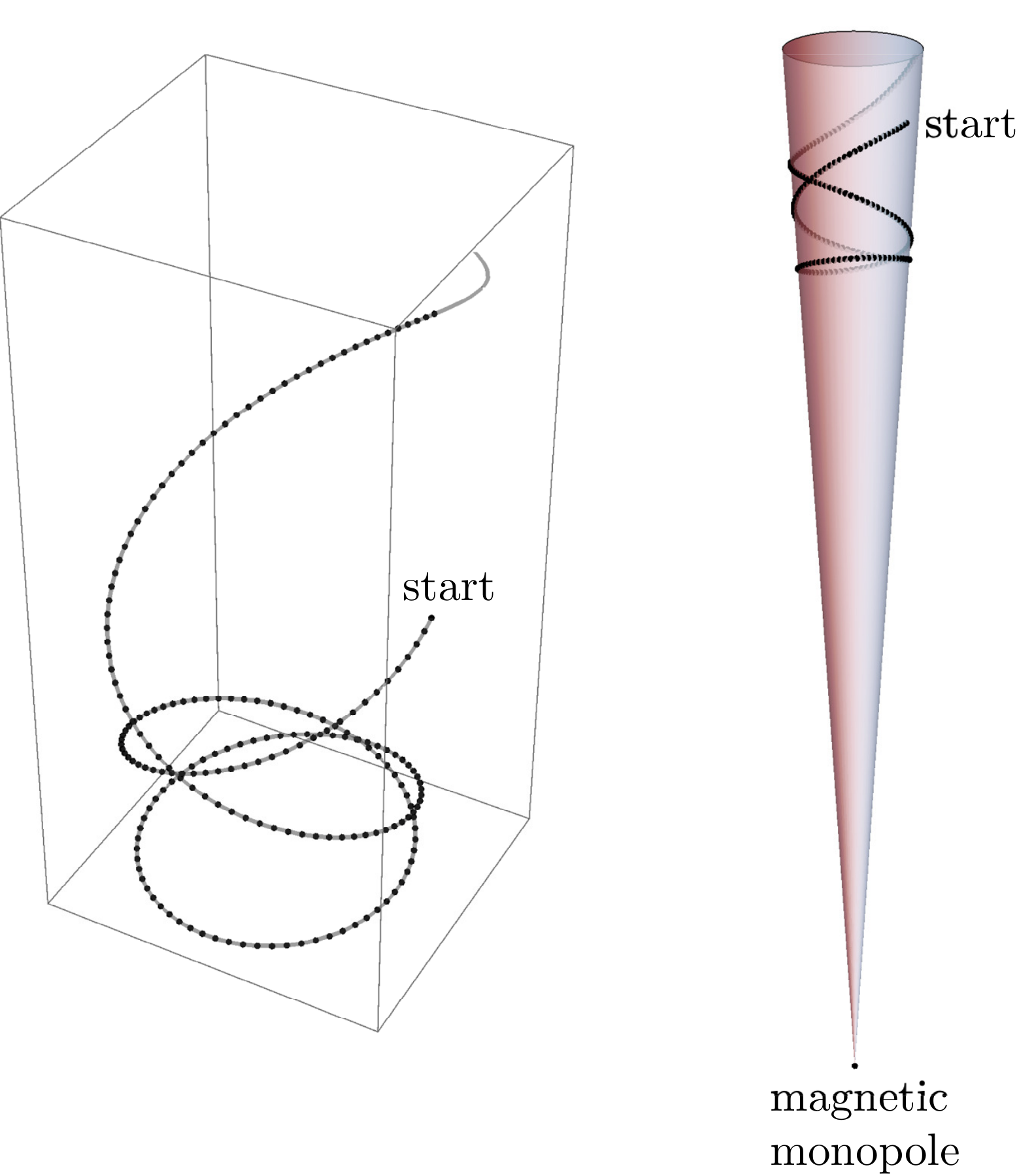} 
   \caption{A particle starts (as shown) moving towards a magnetic monopole along a radially directed field line.  The mirror effect, produced by converging field lines, causes the particle to change direction and head away from the monopole.  On the left, the points are the numerical solution, with the solid gray line the solution from~\refeq{rofphi}.  On the right, we see the cone superimposed with the numerical solution.  The tip of the cone is at the monopole, and it has the opening angle used in the numerical solution, $\theta = 4^\circ$.}
   \label{fig:monopole}
\end{figure}

The numerical solution is shown plotted as points on top of the positions obtained from~\refeq{rofphi} in~\reffig{fig:monopole}.  There we can see the mirror effect as the particle moves down at first, then reverses direction and goes back up.   There is good agreement between the numerical solution and the exact one, with the two overlapping.  The constant ${\bf Q}$ is preserved by the numerical method with norm that varies by only $(\hbox{max}(Q) -\hbox{min}(Q))/(\hbox{min}(Q)) \approx 10^{-7}$ over the portion of the trajectory shown.

Unlike the cases we have considered so far, the component of the motion that circles around a field line is not cylindrical here, even in approximation.  Instead, the charge moves around a cone with constant polar angle $\theta$, with the tip of the cone at the monopole (shown on the right in~\reffig{fig:monopole}), but this change from cylindrical to conical doesn't change the qualitative picture much.  We still have a particle that follows a field line while moving around it, slowing and reversing its direction along the line as it encounters increasing field strength.

\section{Conclusion}

The extension of velocity Verlet to include magnetic fields provides a simple and useful tool for calculating the trajectories of charged particles moving in those fields.  While velocity Verlet is less accurate than some other ODE solving techniques (notably higher-order Runge-Kutta methods), it is easy to implement, and its derivation highlights some of the vector geometry associated with the Lorentz force in the context of Newton's second law.  One can easily extend the method to include additional position-dependent forces.  Going back to~\refeq{LorentzUp}, an additional force $\bar{\bf F}(x)$ would introduce a term like $\bar{\bf F}({\bf x}(t)) + \bar{\bf F}({\bf x}(t + \Delta t))$, where both force evaluations rely on quantities that are known by the time the update is performed, so that these terms would get lumped into the vector ${\bf d}$ in~\refeq{wstart}, at which point the derivation proceeds as in the text.  The method is fast, and can handle multiple particles (making it a favorite of molecular dynamics solvers~\cite{MD}).
We hope this paper serves to increase its use in the undergraduate E\&M curriculum.

\begin{acknowledgments}
The authors thank David Griffiths for useful commentary and for suggesting the monopole magnetic field example.
\end{acknowledgments}


\begin{thebibliography}{0}%
\makeatletter
\providecommand \@ifxundefined [1]{%
 \@ifx{#1\undefined}
}%
\providecommand \@ifnum [1]{%
 \ifnum #1\expandafter \@firstoftwo
 \else \expandafter \@secondoftwo
 \fi
}%
\providecommand \@ifx [1]{%
 \ifx #1\expandafter \@firstoftwo
 \else \expandafter \@secondoftwo
 \fi
}%
\providecommand \natexlab [1]{#1}%
\providecommand \enquote  [1]{``#1''}%
\providecommand \bibnamefont  [1]{#1}%
\providecommand \bibfnamefont [1]{#1}%
\providecommand \citenamefont [1]{#1}%
\providecommand \href@noop [0]{\@secondoftwo}%
\providecommand \href [0]{\begingroup \@sanitize@url \@href}%
\providecommand \@href[1]{\@@startlink{#1}\@@href}%
\providecommand \@@href[1]{\endgroup#1\@@endlink}%
\providecommand \@sanitize@url [0]{\catcode `\\12\catcode `\$12\catcode
  `\&12\catcode `\#12\catcode `\^12\catcode `\_12\catcode `\%12\relax}%
\providecommand \@@startlink[1]{}%
\providecommand \@@endlink[0]{}%
\providecommand \url  [0]{\begingroup\@sanitize@url \@url }%
\providecommand \@url [1]{\endgroup\@href {#1}{\urlprefix }}%
\providecommand \urlprefix  [0]{URL }%
\providecommand \Eprint [0]{\href }%
\providecommand \doibase [0]{http://dx.doi.org/}%
\providecommand \selectlanguage [0]{\@gobble}%
\providecommand \bibinfo  [0]{\@secondoftwo}%
\providecommand \bibfield  [0]{\@secondoftwo}%
\providecommand \translation [1]{[#1]}%
\providecommand \BibitemOpen [0]{}%
\providecommand \bibitemStop [0]{}%
\providecommand \bibitemNoStop [0]{.\EOS\space}%
\providecommand \EOS [0]{\spacefactor3000\relax}%
\providecommand \BibitemShut  [1]{\csname bibitem#1\endcsname}%
\let\auto@bib@innerbib\@empty
\end{thebibliography}%


\begin{thebibliography}{18}
\bibitem{APMOT} Jane M. Repko, Wayne W. Repko, Allan Saaf, ``Charged particle trajectories in simple non-uniform magnetic fields," {\it Am. J. Phys.}, {\bf 59}, 652--655 (1991).
\bibitem{GRIFFITHS} David J. Griffiths, \textsl{Introduction to Electrodynamics}, 4th ed. (Cambridge University Press,  2017).  
\bibitem{VANALLEN} M. Kaan \"Ozt\"urk, ``Trajectories of charged particles trapped in Earth's magnetic field," {\it Am. J. Phys.}, {\bf 80}, 420--428 (2012).
\bibitem{MAPLES} George C. McGuire, ``Using computer algebra to investigate the motion of an electric charge in magnetic and electric dipole fields," {\it Am. J. Phys.}, {\bf 71}, 809--812 (2003).
\bibitem{COMPUTERSTUDIES} Elisha R. Huggins, Jeffrey J. Lelek, ``Motion of electrons in electric and magnetic fields; introductory laboratory and computer studies," {\it Am. J. Phys.}, {\bf 47}, 992--999 (1979).
\bibitem{FRANKLINAJP} J.\ Franklin, K. C.\ Newton, ``Classical and quantum mechanical motion in magnetic fields," {\it Am. J. Phys.}, {\bf 84}(4), 263--269 (2016).
\bibitem{MDVERLET} Q.\ Spreiter, M.\ Walter, ``Classical molecular dynamics simulation with the Velocity Verlet algorithm at strong external magnetic fields," {\it J. Comp. Phys.} {\bf 152}, 1, 102--119 (1999).
\bibitem{JACKSON} John David Jackson, \textsl{ Classical Electrodynamics}, 3rd ed. (Wiley, 1998).
\bibitem{GOLDSTEIN} Herbert Goldstein, Charles P.\ Poole Jr.\ and John L.\ Safko, \textsl{Classical Mechanics}, 3rd ed.  (Pearson, 2001).
\bibitem{VERLET} Loup Verlet, ``Computer `Experiments' on Classical Fluids. I. Thermodynamical Properties of Lennard-Jones Molecules," {\it Phys. Rev.} {\bf 159}(1), 98--103 (1967).
\bibitem{SYMP} Denis Donnelly, ``Symplectic Integrators: An introduction," {\it Am. J. Phys.}, {\bf 73} (10), 938-- 945 (2005).
\bibitem{AT} M. P. Allen, D. J. Tildesley, \textsl{Computer Simulations of Liquids}, (Oxford University Press, New York, 1987).
\bibitem{FRANKLINCMP} Joel Franklin, \textsl{Computational Methods for Physics}, (Cambridge University Press, New York, 2013).
\bibitem{POST}  R. F. Post, ``Summary of UCRL Pyrotron (Mirror Machine) Program," Proceedings of the Second International Conference on Peaceful Uses of Atomic Energy, Geneva, Paper 31, Vol 32, 245--265 (1958).  Available at: 


\texttt{\scriptsize http://www-naweb.iaea.org/napc/physics/2ndgenconf/data/Proceedings\%201958/papers\%20Vol32/Paper31\_Vol32.pdf}



\bibitem{INTNOTE}  We encourage curious readers to find the source current density that produces this field and compare that source with the one that produced the field in~\refeq{Bzonly}.  In addition, one can compare the field in~\refeq{mirrorfield} with the full, off-axis, field produced by a pair of current loops placed at $\pm d$ along the $z$ axis.
\bibitem{PROVIDED} There is also the possibility that the particle is not trapped at all.  One can develop, from the expression for $\mu$, constraints on the ratio of the perpendicular to longitudinal velocity components that prevent trapping.  We leave this interesting opportunity, and its numerical verification, for the reader~\cite{POST, JACKSON}.
\bibitem{MD} E. della Valle, P. Marracino, S. Setti, R. Cadossi, M. Liberti and F. Apollonio, ``Magnetic molecular dynamics simulations with Velocity Verlet algorithm," 2017 XXXIInd General Assembly and Scientific Symposium of the International Union of Radio Science (URSI GASS), Montreal, QC, 2017, pp. 1-4, doi: 10.23919/URSIGASS.2017.8105168.
\end{thebibliography}
\end{document}